\documentclass{ws-p9-75x6-50}

\begin{document}

\title{Scaling Dimensions of Lattice Quantized Gravity}

\author{Herbert W. Hamber}{

\address{Department of Physics and Astronomy \\ University of California \\
Irvine, CA 92697-4575 USA \\
e-mail: hhamber@uci.edu }



\maketitle

\abstracts{
I discuss a model for quantized gravitation based on the simplicial lattice
discretization.
It has been studied in detail using a comprehensive finite
size scaling analysis combined with renormalization group methods.
The results are consistent with a value for the universal 
critical exponent for gravitation $\nu=1/3$, and suggest a simple
relationship between Newton's constant, the gravitational
correlation length and the observable average space-time curvature.
Some perhaps testable phenomenological implications
are discussed, such as the scale dependence of Newton's constant and
properties of quantum curvature fluctuations.}

\section{Introduction}

One of the outstanding problems in physics is
a determination of the quantum-mechanical properties
of Einstein's theory of Gravitation.
Approaches based on linearized perturbation methods have had limited
success so far,
as the underlying theory is known not to be perturbatively renormalizable.
Since gravitational fields are themselves the source for
gravitation already at the classical level,
perturbative results are perhaps of doubtful validity for sufficiently
strong effective couplings.
This is especially true in the quantum domain, where large fluctuations
in the gravitational field appear at short distances.
In general nonperturbative effects can give rise to novel
behavior in quantum field theory, and in
particular to the emergence of non-trivial fixed points of the renormalization
group, a phase transition in statistical mechanics language.
Then the universal 
low energy behavior of field theories is almost completely
determined by the fixed point structure of the renormalization group
trajectories ~\cite{wilson}.

The situation described above clearly bears some resemblance to the theory
of strong interactions, Quantum Chromodynamics, where non-linear
effects are known here to play an important role, and end up restricting
the validity of perturbative calculations to the high energy, short distance
regime ~\cite{gross}.
For low energy properties the lattice formulation, combined with
renormalization group methods and computer simulations, has provided
convincing evidence for quark confinement and chiral
symmetry breaking, two phenomena which are invisible to any order in the
usual perturbative expansion.

\section{The Lattice Theory}

A discrete lattice formulation can
be applied to the problem of quantizing gravitation;
instead of continuous metric fields, one deals with new gravitational
degrees of freedom, the edges, which live only on discrete space-time
points and interact locally with each other.
In Regge's simplicial formulation of quantum gravity ~\cite{regge}
one then approximates
the functional integration over continuous metrics by a discretized
sum over piecewise linear simplicial geometries ~\cite{lesh,hartle}.
In such a model the role of the continuum metric is played
by the edge lengths of the simplices, while local curvature is
described by a set of deficit angles, determined via
known formulae as functions of the given edge lengths.
The simplicial lattice
formulation of gravity is locally gauge invariant ~\cite{gauge} and can be
shown to contain perturbative gravitons in the lattice weak field expansion
\cite{rowi}, making it an attractive and faithful lattice regularization
of the continuum theory.
In the end the original continuum theory of gravity is to be recovered
as the space-time volume is made large and the fundamental lattice
spacing  of the discrete theory is sent to zero, without
having to rely, at least in principle, on any further approximation
to the original continuum theory.

Quantum fluctuations in the underlying geometry are represented
in the discrete theory
by fluctuations in the edge lengths, which can be modeled by
a well-defined, and numerically exact, stochastic process.
In analogy with other field theory models studied by computer,
calculations are usually performed in the Euclidean
imaginary time framework, which is the only formulation amenable
to a controlled numerical study, at least for the immediate foreseeable future.
The Monte-Carlo method, based on the concept of importance sampling,
is well suited for evaluating in a numerically exact way 
the discrete path integral for
gravity and for computing the required averages and correlation
functions.
By a careful and systematic analysis of the lattice results, the
critical exponents can be extracted, and the scaling properties of
invariant correlation functions determined from first principles
\cite{critical}.

The starting point for a non-perturbative study of quantum gravity
is a suitable definition of the discrete Feynman path integral.
In the simplicial lattice approach one starts from the discretized
Euclidean path integral for pure gravity,
with the squared edge lengths taken as fundamental variables,
\be
Z_L = \int_0^\infty \prod_s \; \left ( V_d (s) \right )^{\sigma}
\prod_{ ij } \, dl_{ij}^2 \Theta [l_{ij}^2] 
\exp \left \{ 
- \sum_h \, \Bigl ( \lambda \, V_h - k \, \delta_h A_h 
+ a \, \delta_h^2 A_h^2 / V_h  \Bigr ) \right \}  \;\; .
\label{eq:zlatt} 
\ee
The above expression represents a lattice discretization of the
continuum Euclidean path integral for pure quantum gravity
\be
Z_C = \int \prod_x
\left ( {\textstyle \sqrt{g(x)} \displaystyle} \right )^{\sigma}
\prod_{ \mu \ge \nu } \, d g_{ \mu \nu } (x)
\exp \left \{ 
- \int d^4 x \, \sqrt g \, \Bigl ( \lambda - { k \over 2 } \, R
+ { a \over 4 } \, R_{\mu\nu\rho\sigma} R^{\mu\nu\rho\sigma}
\Bigr ) \right \}  \;\; ,
\label{eq:zcont}
\ee
(with $k^{-1} = 8 \pi G $, and $G$ Newton's constant) and reduces
to it for smooth enough field configurations.
The discrete gravitational measure in $Z_L$
should be considered as the lattice analog of the DeWitt
continuum functional measure.
The $\delta A$ term in the lattice action is the well-known Regge
term ~\cite{regge}, and reduces to the Einstein-Hilbert action
in the lattice continuum limit ~\cite{rowi}.
A cosmological constant term is needed for convergence of the path
integral for large edge lengths,
while the curvature squared term allows one to control the
fluctuations in the curvature ~\cite{lesh}.

In general in the Regge theory 
the correspondence between lattice and continuum operators
is ~\cite{lesh}
\bea
\sqrt{g} \, (x) \; & \to & \; 
\sum_{{\rm hinges} \, h \supset x } \; V_h 
\nonumber \\
\sqrt{g} \, R (x) \; & \to & \; 
2 \sum_{{\rm hinges} \, h \supset x } \; \delta_h A_h
\nonumber \\
\sqrt{g} \, R_{\mu\nu\lambda\sigma} R^{\mu\nu\lambda\sigma} (x) \; & \to & \;
4 \sum_{{\rm hinges} \, h \supset x } \; ( \delta_h A_h )^2 / V_h 
\label{eq:ops}
\eea
where $\delta_h$ is the deficit angle at the hinge (triangle) $h$,
$A_h$ is the area of hinge labelled by $h$, and $V_h$ is the four-volume
associated with the same hinge $h$.  

In practice and for obvious phenomenological reasons one is only interested
in the limit when the higher derivative contributions are small compared
to the rest, $a \rightarrow 0$.
In this limit the theory depends, in the
absence of matter and after a suitable rescaling of the metric, only on
one bare parameter, the dimensionless coupling $k^2 / \lambda $.
Without loss of generality, one can therefore set the bare cosmological
constant $\lambda=1$. In the following discussion
only the case $a=0$ will be
considered.

\section{Phases of Lattice Gravity}

Studies on small lattices suggest a rich scenario
for the ground state of quantum gravity ~\cite{lesh,monte,phases}.
The present evidence indicates that simplicial quantum gravity
in four dimensions exhibits a phase transition (in the bare coupling $G$)
between {\it two phases}:
a strong coupling phase, in which the geometry is smooth at large
scales and quantum fluctuations in the gravitational field eventually
average out and are bounded;
and a weak coupling phase, in which the geometry is degenerate
and space-time collapses into a lower-dimensional manifold, bearing
some physical resemblance to a spiky branched polymer.
Only the smooth, small negative curvature (anti-DeSitter like)
phase appears to be physically acceptable.

Phrased in a different language, the two phases of quantized gravity
can loosely be described as having in one phase
(with bare coupling $G<G_c$, the branched polymer-like phase) ~\cite{phases}
\be
\langle g_{\mu\nu} \rangle \; = \; 0 \;\; ,
\ee
while in the other (with bare coupling $G>G_c$, the smooth phase),
\be
\langle g_{\mu\nu} \rangle \; \approx \; c \; \eta_{\mu\nu} \;\; ,
\ee
with a vanishingly small negative average curvature in the vicinity
of the critical point at $G_c$.

The existence of a phase transition at finite
coupling $G$, usually associated in quantum field theory with the appearance
of an ultraviolet fixed point of the renormalization group, implies
in principle non-trivial, calculable non-perturbative scaling properties
for correlations and effective coupling constants, and in particular
here for Newton's gravitational constant.
Since only the smooth phase with $G>G_c$ has acceptable
physical properties, one would conclude on the basis of fairly general
renormalization group arguments that at least in this lattice model the 
gravitational coupling can only {\it increase} with distance.
In addition the rise of the gravitational coupling in the infrared
region rules out the applicability of perturbation theory
to the low energy domain, to the same extent that such an
approach is deemed to be inapplicable to study the low-energy properties
of asymptotically free gauge theories.

It is a remarkable property of quantum field theories that
a wide variety of physical properties can be determined from a relatively
small set of universal quantities. Namely the
universal leading critical exponents, computed in the vicinity of some
fixed point (or fixed line) of the renormalization group equations.
In the lattice theory the presence of a fixed point or phase transition is
often inferred from the appearance of non-analytic terms in invariant local
averages, such as for example the average curvature defined as
\be
<l^2> { < \int d^4 x \, \sqrt{ g } \, R(x) >
\over < \int d^4 x \, \sqrt{ g } > }
\; \equiv \; {\cal R} (k)
\mathrel{\mathop\sim_{ k \rightarrow k_c}}
- A_{\cal R} \; ( k_c - k ) ^{ 4 \nu - 1 } \;\; .
\ee
From such averages one can infer the value for
$\nu$, the correlation length exponent,
\be
\xi (k) \; \mathrel{\mathop\sim_{ k \rightarrow k_c}} \; A_\xi \;
( k_c - k ) ^{ -\nu } \;\; .
\ee
In terms of the physical correlation length $\xi$ one has
\be
{\cal R} ( \xi ) \; \mathrel{\mathop\sim_{ k \rightarrow k_c}} \;
\xi^{ 1 / \nu - 4 }
\;\; .
\ee
Correct dimensions here can be restored by
supplying appropriate powers of the ultraviolet cutoff, the Planck
length $l_P=\sqrt{G}$.
The fundamental critical 
exponent $\nu$ is related to the derivative of the beta
function for $G$
\be
\beta (G) \, \equiv \, { \partial G \over \partial \log \mu }
\ee
in the vicinity of the ultraviolet fixed point at $G_c$,
\be
\beta ' (G_c) \, = \, - 1/ \nu \;\; .
\ee
Integrating close to the non-trivial fixed point
one obtains for $G > G_c $
\be
m \, = \, \xi^{-1} \, = \,
\Lambda \, \exp \left ( { - \int^G \, {d G' \over \beta (G') } }
\right )
\, \mathrel{\mathop\sim_{G \rightarrow G_c }} \,
\Lambda \, | \, k_c - k |^{ - 1 / \beta ' (G_c) } \;\;\;\; ,
\ee
where $m$ is an integration constant identified with the inverse
correlation length, and $\Lambda=l_P^{-1}$ the ultraviolet cutoff.

This physical correlation length $\xi$ determines the
long-distance decay of the connected invariant two-point correlations at
fixed geodesic distance $d$.
For example
for the curvature correlation one has, for distances much larger compared
to the correlation length $\xi$,
\be
< \sqrt{g} \; R(x) \; \sqrt{g} \; R(y) \; \delta ( | x - y | -d ) >_c \;
\mathrel{\mathop\sim_{d \; \gg \; \xi }} \;\;
d^{- \sigma } \; e^{-d / \xi } \;\; ,
\label{eq:exp}
\ee
while for short distances one expects a power law decay
\be
< \sqrt{g} \; R(x) \; \sqrt{g} \; R(y) \; \delta ( | x - y | -d ) >_c \;
\mathrel{\mathop\sim_{d \; \ll \; \xi }} \;\; 
{1 \over d^{\; 2 \; (4-1/ \nu)} } \;\; .
\label{eq:pow1}
\ee
One further result is that according to
the renormalization group the scale dependence of the
effective Newton constant is given by
\be
G(r) \; = \; G(0) \left [ \, 1 \, + \, c \, ( r / \xi )^{1 / \nu} \, 
+ \, O (( r / \xi )^{2 / \nu} ) \, \right ] \;\; ,
\label{eq:grun}
\ee
with $c$ a calculable numerical constant.
Here the scale $\xi^{-1}$ plays a role very
similar to the
scaling violation parameter $\Lambda_{\overline{MS}}$ of QCD.
It seems natural, although paradoxical at first, to associate $\xi$ with
some macroscopic cosmological length scale.

It should be clear from this brief discussion that the
critical exponents provide a wealth 
of useful information about the continuum theory.
In reality, the complexity of the lattice interactions and the practical
need to sample
many statistically independent field configurations contributing to the
path integral leads to the requirement of powerful computational
resources.
The results discussed here ~\cite{critical} were obtained using
a dedicated custom-built 20-GFlop 64-processor parallel
supercomputer, described in detail elsewhere ~\cite{aeneas}.

Table I ~\cite{critical} summarizes the results obtained for the
critical point
$k_c =1/8 \pi G_c$ and the critical exponent $\nu$.
From the numerical calculations one finds
\be
k_c = 0.0636(11) \;\;\;\;\;  \nu = 0.335(9) \;\;\;\; ,
\ee
which clearly suggests $\nu = 1/3$ for pure quantum gravity.

\begin{table}

\begin{center}
\begin{tabular}{|l|l|l|}
\hline\hline
Method  & $k_c$ & $\nu$ 
\\ \hline \hline
${\cal R}$ vs. $k$ & 0.0630(11) & 0.330(6)
\\ \hline
${\cal R}^{3}$ vs. $k$ & 0.0639(10) & -
\\ \hline
$\chi_{\cal R}$ vs. $k$ & 0.0636(30) & 0.317(38)
\\ \hline
$\chi_{\cal R}^{3/2}$ vs. $k$ & 0.0641(17) & -
\\ \hline
$\chi_{\cal R}/(\langle l^2 \rangle {\cal R})$ vs. $k$ & 0.0635(11) & 0.339(9)
\\ \hline
$\chi_{\cal R}$ vs. ${\cal R}$ & - & 0.328(6)
\\ \hline \hline
${\cal R}$ FS scaling & - & 0.333(2)
\\ \hline
$\chi_{\cal R}$ FS scaling & - & 0.318(10)
\\ \hline \hline
\end{tabular}
\end{center}
\label{grav}

\center{\small {\it
Table I: Summary table for the critical point $k_c$ and the critical
exponent $\nu$,
as obtained from lattices with up to $16^4$ sites.
The last three entries assume a critical point at $k_c = 0.0636$.
\medskip}}


\end{table}
\vskip 10pt

\section{Critical Exponents and Phenomenology}

This section contains a brief discussion of some of the consequences that
follow from the results presented in the previous section, and
in particular the result $\nu=1/3$.

First let us notice that the value $\nu=1/3$ does not correspond to any
known field theory or statistical mechanics model in four
dimensions.
Indeed for all scalar field theories (spin $s=0$) in four
dimensions it is known that $\nu=1/2$, while for the compact Abelian U(1)
gauge theory ($s=1$) one has $\nu=2/5$.
The value $\nu=1/3$ for $s=2$
is then consistent with the simple interpolation $\nu^{-1} = 2+s/2$.

One distinctive feature of the results is the appearance of a
gravitational correlation length $\xi$.
Naively one would expect, simply on the basis of dimensional
arguments, that the curvature scale should get determined by this correlation
length
\be
{\cal R} \; \mathrel{\mathop\sim_{ {\cal R} \rightarrow 0}} \; 1/ \xi^2 \;\; ,
\label{eq:naive}
\ee
but one cannot in general exclude the appearance of some non-trivial exponent.
In the previous section arguments have been given in support
of the value $\nu=1/3$ for pure gravity. 
From the equation relating the average curvature to the gravitational
correlation length one has
\be
{\cal R} ( \xi ) \; \mathrel{\mathop\sim_{ k \rightarrow k_c}} \;
{ 1 \over l_P^{2-d+1/\nu} \xi^{d-1/\nu} } \;\; .
\label{eq:rm1}
\ee
Here the correct dimension for the average curvature ${\cal R}$
has been restored by supplying
appropriate powers of the ultraviolet cutoff, the Planck length $l_P=\sqrt{G}$.
For $\nu=1/3$ in four dimensions
one then obtains the remarkably simple result
\be
{\cal R} ( \xi ) \; \mathrel{\mathop\sim_{ k \rightarrow k_c}} \;
{ 1 \over l_P \; \xi } \;\;\;\; .
\label{eq:rm2}
\ee
An equivalent form can be given in terms of the curvature scale $H_0$,
defined through $R=-12 H_0^2$ which has dimensions of a mass
squared. Then close to the critical point
\be
H_0^2 \; = \; C_H \; \mu_P \; m \;\; ,
\label{eq:hub}
\ee
where $\mu_P = 1/\sqrt{G}$ is the Planck mass, $m=1/\xi$ is the
inverse gravitational correlation length, and $C_H \approx 4.9$ a numerical
constant of order one (the value for $C$ is extracted from the
known numerical values for ${\cal R}$ and $m$ close to the critical
point at $k_c$). It is amusing to note that this result is
reminiscent of the pcac relation in pion physics.

One can raise the legitimate concern of how these results are changed
by quantum fluctuations of matter fields.
In the presence of matter fields coupled to gravity
one expects the value for $\nu$
to change due to vacuum polarization loops containing these fields.
But based on general arguments, one would expect fields whose masses are
significantly above $m=1/\xi$ to give negligible contributions to
vacuum polarization loops, and thus leave the universal
critical exponents which characterize the large distance behavior unaffected.

It seems natural to identify $H_0$ with either some (negative) average
spatial curvature, or possibly with the Hubble constant determining the
macroscopic expansion rate of the present universe ~\cite{phases}.
In the Friedmann-Robertson-Walker model of standard cosmology
on has for the Ricci scalar
\be
R_{Ricci} \; = \; -6 \left \{  \left ( { \dot{R} \over R } \right )^2 \; + \;
{ k \over R^2 } \; + \; { \ddot{R} \over R }  \right \} \;\;\; ,
\label{eq:ricci} 
\ee
where $R(t)$ is the FRW scale factor, and $k=0,\pm 1$ for spatially
flat, open or closed universes respectively.
Today the Hubble constant is given by $H_0^2 = ( \dot{R} / R )_{t_0}^2 $,
but it is eventually expected to show some slow variation in time.
Its characteristic length scale today $c H_0^{-1} \approx 10^{28} cm$
is comparable to the extent of the visible universe.
Under such circumstances one would expect
the gravitational correlation $\xi$ to be significantly larger than
$c H_0^{-1}$.

A potential problem arises though in trying to establish a relationship
between quantities which are truly constants (such as the ones appearing
in Eq.~(\ref{eq:hub})), and $H_0$ which most likely depends on time
(the only exception being the steady state cosmological
models, where $H$ is truly a constant of nature;
these models are not favored by present observations, including
detailed features of the cosmic background radiation).
In any case it is clear that some of these considerations
are in fact quite general, to the extent that they rely on general
principles of the renormalization group and are not tied to any
particular value of $\nu$, although the favored value $\nu=1/3$
clearly has some aesthetic appeal.

One further observation can be made regarding the running of $G$
in Eq.~(\ref{eq:grun}).
Assuming the existence of an ultraviolet fixed point,
the effective gravitational coupling is given 
for ``short distances'' $r \ll \xi$ by
\be
G(r) \; = \; G(0) \left [ \, 1 \, + \, c \, ( r / \xi )^{3} \, 
+ \, O (( r / \xi )^{6} ) \, \right ] \;\; ,
\label{eq:run1} 
\ee
with $c$ a calculable numerical constant of order one.
The appearance of $\xi$ in this equation, which is a very large
quantity by Eq.~(\ref{eq:hub}), suggests that the leading
scale-dependent correction which gradually increases the strength of the
effective gravitational interaction as one goes to larger and
larger length scales, should be exceedingly small.
It also suggests that the deviations from classical general
relativistic behavior for most physical quantities is in the end
practically negligible.

It is only for distances comparable to or larger than $\xi$ that the
gravitational potential starts to weaken and fall off exponentially,
with a range given by the gravitational correlation length $\xi$,
\be
V(r) \; \mathrel{\mathop\sim_{ r \; \gg \; \xi }} - \; G(r) \;
{ \mu_1 \mu_2 \;  e^{-r/\xi } \over r } \;\; .
\label{eq:potexp}
\ee
In many ways these results appear qualitatively consistent with the expected
behavior of
the tree-level graviton propagator in anti-de Sitter space. 
In the real world the range $\xi$ must be of course very large; from the fact
that super-clusters of galaxies apparently do form, one can set an
observational lower limit $ \xi > 10^{25} cm$.

It is unclear to what extent gravitational correlations can be 
measured directly.
From the definition of the curvature correlation function 
in Eq.~(\ref{eq:pow1}) one has for ``short distances'' $r \ll \xi$
and for the specific value $\nu=1/3$ the remarkably simple result
\be
< \sqrt{g} \; R(x) \; \sqrt{g} \; R(y) \; \delta ( | x - y | -d ) >_c \;
\mathrel{\mathop\sim_{d \; \ll \; \xi }} \;\; 
{A \over d^2 } \;\;\; ,
\label{eq:pow2}
\ee
with $A$ a calculable numerical constant of order one.
One should contrast this behavior with the semiclassical result attained
close to two dimensions (and which incidentally coincides with the
lowest order weak field expansion result), which gives instead
for the power the value $ 2(d-1/\nu) \sim 2 (d - (d-2)) \sim 4 $, as
expected on the basis of naive dimensional arguments ($R \sim \partial^2 h$).

Next one can look at local curvature fluctuation.
If one considers the curvature $R$ averaged over a spherical volume
$V_r = 4 \pi r^3 /3$,
\be
\overline{ \sqrt{g} \; R } \; = \; { 1 \over V_r } \;
\int_{V_r} d^3 \vec{x} \;  \sqrt{g(\vec{x},t)} \; R(\vec{x},t)
\ee
one can compute the corresponding variance in the curvature
\be
\left [ \delta ( \sqrt{g} \; R ) \right ]^2 \; = \; { 1 \over V_r^2 } \;
\int_{V_r} d^3 \vec{x} \; \int_{V_r} d^3 \vec{y} \;
< \sqrt{g} \; R(\vec{x}) \; \sqrt{g} \; R(\vec{y}) >_c \; = \; 
{9 A \over 4 \; r^2 } \;\;\; .
\ee
As a result the r.m.s. fluctuation of $\sqrt{g} R$ averaged over a 
spherical region of size $r$ is given by
\be
\delta ( \sqrt{g} \; R ) \; = \; 
{ 3 \sqrt{A} \over 2 } \; {1 \over r } \;\;\; ,
\ee
while the Fourier transform power spectrum at small $\vec{k}$ is given by
\be
P_{\vec{k}} \; = \; \vert \; \sqrt{g} \; R_{\vec{k}} \; \vert^2 \; = \; 
{ 4 \pi^2 A \over 2 V } \; {1 \over k } \;\;\; .
\ee
One can use Einstein's equations to relate the local curvature to the
(primordial) mass density. From the Einstein field equations
\be
R_{\mu\nu} -
{\textstyle {1 \over 2} \displaystyle}
g_{\mu\nu} R \; = \; 8 \pi T_{\mu\nu} \;\;
\label{eq:ein}
\ee
for a perfect fluid
\be
T_{\mu\nu} \; = \; p g_{\mu\nu} + (p+\rho) u_\mu u_\nu \;\;
\label{eq:fluid}
\ee
one obtains for the Ricci scalar, in the limit of negligible pressure,
\be
R (x) \; \approx \; 8 \pi G \; \rho (x) \;\;\; .
\ee
As a result one expects for the density fluctuations a power law decay
of the form
\be
< \rho (x) \; \rho (y) >_c \;
\mathrel{\mathop\sim_{ |x-y| \; \ll \; \xi }} \;
{1 \over |x-y|^2 } \;\; .
\label{eq:pow3}
\ee
Similar density correlations have been estimated from observational data
by analyzing known galaxy number density distributions, giving a
value for the exponent of about $1.77 \pm 0.04$ for distances in the $10kpc$
to $10Mpc$ range.

\section{Outlook}

Numerical simulation methods combined with
modern renormalization group arguments
can provide detailed information on non-perturbative aspects of
a lattice model of quantum gravity.
One then finds that the lattice model has two phases,
only one of which is physically acceptable. 
In spite of the fact that the Euclidean theory becomes unstable as one
approaches the critical point at $k_c$, it is still possible to
determine by a straightforward analytic continuation the physical properties
of the model in the vicinity of the true fixed point,
defined as the point where a non-analiticity develops in the 
strong coupling branch of $Z_L(k)$. There scaling implies that
the physical correlation $\xi$ diverges.

If this prescription is followed, an estimate for the non-perturbative
Callan-Symanzik beta function in the vicinity of the fixed point
can be obtained, to leading order in the deviation of the bare coupling
from its critical value.
The resulting scale evolution for the gravitational constant is then
quantitatively quite small, if one assumes that the scaling violation
parameter is related to an average curvature and its characteristic
scale $H_0$.
Its infrared growth, consistent with the general idea that gravitational
vacuum polarization effects cannot exert any screening, suggests
that low energy properties of quantum gravity are inaccessible by
weak coupling perturbation theory: low energy quantum gravity
is a strongly coupled theory.
As pointed out in the discussion there are
a number of attractive features to the pure gravity result $\nu =1/3$,
including a simple form for the curvature correlations at short distances. 

It seems legitimate to ask the question whether the present lattice model for
quantum gravity provides any insight into the problem of the cosmological
constant. The answer is both yes and no. 
To the extent that a naive prediction of quantum gravity is that
the curvature scale should be of the same order of the Planck 
length, ${\cal R} \sim 1/G$, the answer is definitely yes.
Indeed it can be regarded as a non-trivial result of the lattice models for
gravity that a region in coupling constant space can be found where
space-time is stiff and 
the curvature can be made much smaller than $1/G$. In fact the evidence
indicates that the average curvature ${\cal R}$ vanishes at the critical
point $k_c$.
And this is achieved with a bare cosmological constant $\lambda$ which
is of order one in units of the cutoff.
Phrased differently, the dimensionless ratio between the renormalized
and the bare cosmological constant becomes arbitrarily small towards
the critical point.

At the same time the effective long distance cosmological constant is
clearly non-vanishing and of order $1/\xi$ as a consequence of dimensional
transmutation. The value zero is only obtained
when $\xi$ is exactly infinite,
which happens only at the critical point $k_c$. 
Thus to make the effective cosmological constant small requires a fine
tuning, in the sense that the bare coupling $k_c-k$ has to be small.
But since the correlation length determines the corrections to the Newtonian
potential (and in particular its eventual decrease for large enough
distances), it would seem unnatural to have a short correlation
length $\xi$: in such a world there would be no long-range gravitational
forces, and separate space-time domains would have decoupled fluctuations.
From this perspective, long range forces and a small cosmological
constant go hand in hand. Quantum fluctuation effects show that
hyperbolic space-times with small curvature radii
cannot sustain long-range gravitational forces, at least in this lattice model.

\vskip-\lastskip
\pagebreak

\section{Acknowledgments}

The {\it Aeneas} Supercomputer Project is supported in part by
the Department of Energy, the National Science Foundation and
the University of California.

\end{document}